\documentclass[11pt,prd,showpacs,onecolumn,preprintnumbers
]{revtex4}

\usepackage{amsmath}  \usepackage{amsfonts} \usepackage{psfrag}
\usepackage{graphicx}

\newcommand{\df}{\dot{f}}\newcommand{\da}{\dot{a}}\newcommand{\dal}{\dot{\alpha}}

\newcommand{\dpsi}{\dot{\psi}}
\newcommand{\Lc}{L_{NM}}
\newcommand{\ep}{\epsilon}  
\newcommand{\Ne}{\mathrm{N}_{e\mathrm{-folds}}}

\begin{document}
\title{Cosmology with non-minimally coupled Yang--Mills field}
\author{Evgeny A. Davydov} \email{eugene00@mail.ru} \affiliation{Bogoliubov
Laboratory of Theoretical Physics, JINR, 141980, Dubna, Moscow
region, Russia}
\author{Dmitry V. Gal'tsov}\email{galtsov@phys.msu.ru}
\affiliation{Department of Theoretical Physics, Moscow State
University, 119899, Moscow, Russia}

\begin{abstract}
We discuss cosmological model with homogeneous and isotropic
Yang-Mills field non-minimally coupled to gravity through an
effective mass term. In this model conformal symmetry is violated
which leads to possibility of inflationary expansion. Parameters of
non-minimal coupling have relatively ``natural'' values in the
regime of sufficiently long acceleration stage.
\end{abstract}
\maketitle
\section{Introduction}
Vector fields were suggested in cosmology as alternative to scalar
fields as inflation and dark energy drivers
\cite{Ford,Arm,Kiselev:2004py,Wei:2006tn}. Apart from being
physically appealing as well-understood and common component of all
existing theories,they may be useful, in particular, in generating
appropriate curvature perturbations
\cite{Dimopoulos:2009am,Dimopoulos:2011ws,Karciauskas:2011fp}.
Especially attractive seem to be non-Abelian models, which admit
magnetic type  field configurations compatible with isotropy and
homogeneity of space-time
\cite{Galtsov:1991un,VoGa,Moniz:1990hf,Darian:1996mb} and may
therefore be used in the standard Friedmann-Robertson-Walker setting
without averaging needed in the case of vector singlet to avoid
anisotropy. If their dynamics is ruled by the standard Yang-Mills
(YM) conformally invariant action, the corresponding equation of
state is that of the hot universe \cite{Galtsov:1991un}, but
violation of conformal symmetry may lead to different equations of
state. In particular, the Born-Infeld lagrangian generates an
equation of state which interpolates between the hot regime at low
densities and a zero acceleration regime at high density
\cite{BI,elizalde,Fuzfa:2005qn,Fuzfa:2006}. Stronger violation of
conformal symmetry in the purely YM sector may produce accelerating
expansion of inflationary dark energy type \cite{1} exhibiting
typically a finite acceleration period. Various realizations of this
scenario were suggested \cite{Zhao:2005bu}--\cite{Shchigolev:2011kr}
with different motivation. Cosmological perturbations in the YM
cosmology were studied in \cite{pert,zraa2009,kmm}. It is also worth
noting that if one treats inflation as associated with the Higgs
sector of some gauge theory, the excitation of the Yang-Mills
component becomes inevitable and modifies inflationary regime in a
non-trivial way
\cite{Gal'tsov:2010dd,Galtsov:2011aa,Davydov:2011aa}.

Here we would like to investigate cosmological dynamics of
Yang-Mills filed non-minimally coupled to gravity via
curvature-dependent mass term. Earlier proposals of non-minimally
coupled Yang-Mills in cosmology include
\cite{Balakin:2007mp,Bamba:2008xa,Banijamali:2011ep,Elizalde:2010xq},
but our approach here is different. To introduce non-minimal
coupling one usually probe various combination the Riemann tensor
with the field strength and the vector potential. In the latter case
the gauge invariance of the theory is destroyed at  classical level,
but one consider this just as the simplified for cosmological
purposes effect of spontaneous symmetry breaking mechanism of the
gauge theory giving mass to $W$-boson.
\section{The model}
We introduce  the non-minimal coupling of the YM potential
$A^{a}_\mu$ to the Ricci tensor generically depending on two
constants together with the ordinary mass term. This form is
motivated by absence of higher derivatives in the resulting
equations, so the theory seems to be ghost-free. With this in mind
we choose as the lagrangian the sum of the Einstein-Hilbert term,
the standard YM term and the non-minimal term $L=L_{EH}+L_{YM}+\Lc$,
namely:
\begin{equation}\label{dL}
L=-\frac{ {M}_{Pl}^2}{2}R-\frac{1}{4}
F^{a}_{\mu\nu}F^{a\mu\nu}+\left(\nu R/4-m^2/2\right) A^{a}_\mu
A^{a\mu}+(\lambda-\nu) S_{\mu\nu}A^{a\mu} A^{a\nu}\,.
\end{equation}
Here $ {M}_{Pl}=1/\sqrt{8\pi G}$ is the modified Plank mass, and
$\nu,\lambda$ are new \emph{dimensionless} parameters and $m^2$ is
the mass term which looks natural within such a model. The Ricci
coupling is split into the scalar curvature coupling and coupling to
Schouten tensor, defined in $D$ dimensions as
\begin{equation}\label{Schouten}
    S_{\mu\nu}=\frac{1}{D-2}\left(R_{\mu\nu}-\frac{R}{2(D-1)}g_{\mu
    \nu}\right)\,,
\end{equation}
its advantage will be clear later.

It is convenient to choose the  length scale $l=(1/e {M}_{Pl})$ and
replace the mass by the dimensionless parameter $\mu=(m/e
{M}_{Pl})^2$. The  interval in dimensionless coordinates reads
\begin{equation}\label{metric}
     ds^2=l^2\left\{-N^2 dt^2+a^2\left[d\chi^2+\Sigma^{2}_{k}(\chi)
 (d\theta^2+\sin^2\theta d\varphi^2)\right]\right\}\,,
\end{equation}
where $\Sigma_{k}(\chi)=\{\sin\chi,\chi,\sinh\chi\}$ for the
closed, flat and open universe, labeled by $k=1,0,-1$,
correspondingly.

The scalar curvature and the Schouten tensor contain the second
order derivatives
\begin{eqnarray}\label{Ricci}
    &&R = 6\left[(\da/aN)\dot{}/N+2\da^2/a^2
    N^2+k/a^2\right];\\
   && S^{0}_0 = (\da/aN)\dot{}/N+\da^2/2a^2 N^2-k/2a^2,
   \quad S^{i}_{i}=\da^2/2a^2N^2+k/2a^2\;.
\end{eqnarray}
In General Relativity it is common to integrate by parts separating
the total derivative
\begin{equation}
    \frac12 R\sqrt{-g}=3\left[a^3(\da/aN)\dot{}+2a\da^2/N+
    ka N\right]=3\left[ka N-a\da^2/N\right]+\mbox{div}\,.
\end{equation}

The ansatz for the YM potential preserving the isotropy and
homogeneity was constructed in \cite{Galtsov:1991un} for all $k$. In
the temporal gauge $A^{a}_0=0$ we have
\begin{equation}\label{AA}
A^{a}_i A^{a j}=\delta^{j}_{i}\frac{(k-f)^2}{a^2}\,,\quad -\frac14
F_{\mu\nu}^a F^{\mu\nu}_a=\frac{3\df^2}{2a^2N^2}-
    \frac{3(k-f^2)^2}{2a^4}\,.
\end{equation}

The same as for the curvature term, one can shift the second
derivative in the scalar curvature coupling term to the gauge field:
\begin{equation}\label{RAA}
    \frac{1}{4} R A^{a}_\mu A^{a\mu}\sqrt{-g}=3\df(k-f)\da/N+
    \frac32 (k-f)^2\left[kN/a+\da^2/aN\right]+\mbox{div}\,.
\end{equation}
Note that the term $3\df(k-f)\da/N$ looks quite similar to the
topological term
$$\varepsilon^{\mu\nu
\alpha\beta}F^{a}_{\mu\nu}F^{a}_{\alpha\beta}\sqrt{-g}
=3\df(k-f^2)\,.$$  It has the form of the interaction with an
``axion'' $\da/N$ --- which is a common trick to make the
topological term to contribute into dynamics.

The coupling of the YM field amplitude to the Schouten tensor does
not contain the second derivative terms:
\begin{equation}\label{SAA}
    S_{\mu\nu}A^{a\mu}
    A^{a\nu}\sqrt{-g}=\frac{3}{2}(k-f)^2\left[kN/a+\da^2/aN\right]\,.
\end{equation}

Collecting all the terms together and omitting the factor $3$, one
obtains the following effective one-dimensional Lagrangian:
\begin{equation}\label{L1}
\begin{split}
    &L_{\mathrm{eff}}=ka
    N-\frac{a\da^2}{N}+\frac{a\df^2}{2N}-(k-f^2)^2
    \frac{N}{2a}+\Lc,\quad\mbox{where}\\
    &\Lc=-\frac{\mu
    }{2}(k-f)^2 aN+\nu\df(k-f)\frac{\da}{N}+\frac{\lambda}{2}(k-f)^2\left[\frac{kN}{a}+\frac{\da^2}{aN}\right].
    \end{split}
\end{equation}
So there arise three different dynamical terms in $\Lc$ due to
non-minimal coupling, which can be switched off by making zero the
corresponding coupling constants.

\section{Slow-roll} Our goal is to describe the slow-roll inflation within the
above model, so we will make some preparations to make future
analysis a bit easier.  First, we omit the contribution of the
curvature term during the phase of the fast expansion,  though it
contains a very interesting mode of the cosmological sphaleron  in
the case  $k=1$ due to the YM self-interaction potential
\cite{Gist,Volkov:1993gp}, which will be discussed elsewhere. So, in
the what follows, $k=0$.

Next, for the YM mode the conformal field amplitude $\psi=f/a$ is
quite natural, while the metric variable can be chosen in the
exponential form $a=\exp\alpha$. This allows us to simplify the
contribution of the metric into the kinetic term, so  the full
Lagrangian now takes the form:
\begin{equation}\label{L2}
\begin{split}
    &L=\frac{e^{3\alpha}}{2N}\left[-A(\psi)\dal^2+2B(\psi)\dal\dpsi
    +\dpsi^2\right]-N e^{3\alpha}V(\psi), \quad\mbox{where}\\
&A(\psi)=1+2\nu\psi-(\lambda+1)\psi^2,\quad
B(\psi)=(1-\nu)\psi,\quad V(\psi)=\frac{1}{2}(\mu\psi^2+\psi^4)\,.
    \end{split}
\end{equation}
The above dynamical system has the constraint, obtained by the
variation with respect to $N$, which in the form of the Friedman
equation reads:
\begin{equation}\label{constr1}
\frac{\dal^2}{2}A(\psi)=\dal\dpsi
    B(\psi)+\frac{\dpsi^2}{2}+N^2 V(\psi)\,.
\end{equation}

The variation with respect to $\alpha$ and $\psi$ gives the
equations of motion:
\begin{eqnarray}
  &&[N^{-1}e^{3\alpha}(B\dpsi-A\dal)]\,\dot{}
     = \frac{3 e^{3\alpha}}{2N}\left[-A\dal^2+2B\dal\dpsi
    +\dpsi^2-2N^2 V\right], \label{eq_t1}\\
  &&[N^{-1}e^{3\alpha}(\dpsi+B\dal)]\,\dot{} =
    \frac{ e^{3\alpha}}{2N}\left[-A_\psi\dal^2+2B_\psi\dal\dpsi-2N^2V_\psi\right]\,. \label{eq_t2}
\end{eqnarray}
Here $A_\psi,B_\psi,V_\psi$ denote the partial derivative with
respect to $\psi$. Instead of fixing the time gauge we may proceed
with an invariant description, choosing the $\alpha$ as an
independent variable. Introducing the Hubble parameter, $H\equiv
d\alpha/N dt$, we may rewrite the system as
\begin{eqnarray}
  &&H[H e^{3\alpha}(B\psi'-A)]'
     = -\frac32 H^2 e^{3\alpha}\left[A-2B\psi'
    -\psi'^2+2H^{-2}V\right], \label{eq_a1}\\
  &&H\left[H e^{3\alpha}(\psi'+B)\right]' =
    -\frac12 H^2 e^{3\alpha}\left[A_\psi-2B_\psi\psi'+2H^{-2}V_\psi\right]\,,
     \label{eq_a2}
\end{eqnarray}
where  prime denotes the derivative with respect to $\alpha$. Using
the constraint, the potential term can be expressed as
\begin{equation}\label{constraintV}
   2H^{-2} V=A-2B\psi'
    -\psi'^2,
\end{equation}
and used in the above equations.

Now let us introduce the slow-roll parameters  which are usually
used to detect the inflationary stage in the dynamics of the system:
\begin{equation}\label{slowroll}
    \ep=-\frac{\dot{H}}{H^2N}=-\frac{H'}{H},\quad \delta=-\frac{\dpsi}{H\psi
    N}=-\frac{\psi'}{\psi}\,.
\end{equation}
Of course they are also independent on the choice of gauge. We may
rewrite the system of equations~(\ref{eq_a1}--\ref{eq_a2}),
replacing $\psi',\psi'',H'$ by $\delta,\delta',\ep$ and using the
constraint to avoid appearance of the $H^{-2}$ term:
\begin{eqnarray}
  && (\ep-3)(B\psi\delta+A)+(B_\psi\psi+B)\psi\delta^2-B\psi\delta'=
   -3(A+2B\psi\delta -\psi^2\delta^2)\,, \\
 &&(\ep-3)(\psi\delta-B)+\psi(\delta^2-\delta')-B_\psi\psi\delta
  =\nonumber\\&& -\frac{1}{2}(A_\psi+2B_\psi\psi\delta)-\frac{1}{2}[A+2B\psi\delta
  -\psi^2\delta^2](\ln V)_\psi\,.
\end{eqnarray}
Finally collect terms with different powers of $\ep,\delta$:
\begin{eqnarray}
  &&A\ep+3B\psi\delta=B\psi\delta'-(B_\psi\psi+B-3\psi)\psi\delta^2-B\psi\delta\ep, \\
 && 3B+\frac12 \left[A_\psi+A(\ln V)_\psi\right]+\left[B(\ln V)_\psi-3\right]\psi\delta-B\ep
  = \psi^2\delta'+\left[\frac{\psi}{2}(\ln V)_\psi-1\right]\psi\delta^2-\psi\delta\ep.
\end{eqnarray}
Since the constraint was already incorporated in these two
equations, they provide the independent conditions on the slow-roll
parameters. The first equation gives:
\begin{equation}\label{epdelta}
    \ep=-(3B\psi/A)\delta+O(\delta^2).
\end{equation}
Then the second equation implies the relation on the initial state
of the system:
\begin{equation}\label{istate}
    \delta+\frac{A[A(\ln V)_\psi+A_\psi+6B]}{2\psi[AB(\ln
    V)_\psi-3A+3B^2]}=O(\delta^2,\delta').
\end{equation}
If the initial conditions $\{\psi_i,\psi_i'\}$ ensure both
$\delta_i=-\psi_i'/\psi_i\ll 1$ and the l.h.s. of the
relation~(\ref{istate}) to vanish, then one has $\delta'\sim
O(\delta^2),\,\ep\sim\delta$ which signals a slow-roll regime. Now
assume for simplicity that neither the quantity $3B\psi/A$ itself,
nor its derivative with respect to $\psi$ is singular in the
corresponding region of the phase space, to ensure that $\ep'\sim
O(\delta^2)$ as well. Also one has to ensure that the value $H^2$
from the constraint~(\ref{constraintV}) is positive for the chosen
initial conditions: this is the additional restriction, which was
not taken into account before.

To estimate the number of $e$-folds gained by the scale factor
during the slow-roll inflation, one can just treat the
expression~(\ref{istate}) as the function of $\psi:\;\delta(\psi)$.
Then by definition $d\alpha=-[\psi\delta(\psi)]^{-1}d\psi$ and
\begin{equation}
    \Ne=\alpha_e-\alpha_i=-\int_{\psi_i}^{\psi_e}\frac{d\psi}{\psi\delta(\psi)},
\end{equation}
where the exit from the slow-roll may occur when
$|\delta(\psi_e)|=1$. The full expression is rather complicated,
yet one can find that in the trans-Planckian region, $|\psi_i|\gg
1$, $\psi^2\gg|\mu|$,
\begin{equation}\label{delta}
    \delta(\psi)\approx
    -\frac{[2\nu-\psi(\lambda+1)][5\nu-3\psi(\lambda+\nu)]}{(3\nu^2+4\lambda\nu-\lambda-2\nu+2)\psi^2}.
\end{equation}
In addition, the constraint~(\ref{constraintV}) will provide
$H^2>0$ if $A+2B\psi\delta-\psi^2\delta^2>0$.

For the convenient choice of the parameters, say, $\lambda=-1$ or
$\lambda=-\nu$  one can easily get the answer. Indeed, one has now
$\delta\sim 1/\psi$ and $\Ne\sim\psi_i$. Of course, the value
$\Ne$ is calculated up to  the factor of unity, since the exit
condition, $\delta_e\sim 1$, has just the same level of accuracy.
And in general case the analytic analysis is quite complicated. So
we now better proceed with a brief numerical calculations.

\section{Numeric analysis and Outlook}

It is convenient to solve numerically the
system~(\ref{eq_t1}--\ref{eq_t2}) in the gauge $N=1$. Mention that
the system of equations $M_{ij}(q)\ddot{q}^j=\Phi_{i}(q,\dot{q})$
has a singularity  when $\Delta=\det(M)$ vanishes. In our case the
corresponding determinant is $\Delta\sim B^2(\psi)+A(\psi)$. But the
slow-roll initial state automatically ensures  that the regime is
non-singular (the dynamics can not be `slow' in the vicinity of
singularity).

\begin{figure}[p]
\hbox to\linewidth{\hss%
\psfrag{a}{\LARGE{$\psi_i=100$}}\psfrag{b}{\LARGE{$\psi_i=10$}}
\psfrag{c}{\LARGE{$\psi_i=1$}} \psfrag{d}{\LARGE{$\psi_i=0.1$}}
\psfrag{x}{\huge{$\mbox{Abscissae}:\:-5<\nu<5$}}
\psfrag{y}{\huge{$\mbox{Ordinates}:\:-5<\lambda<5$}}

    \resizebox{7cm}{7cm}{\includegraphics{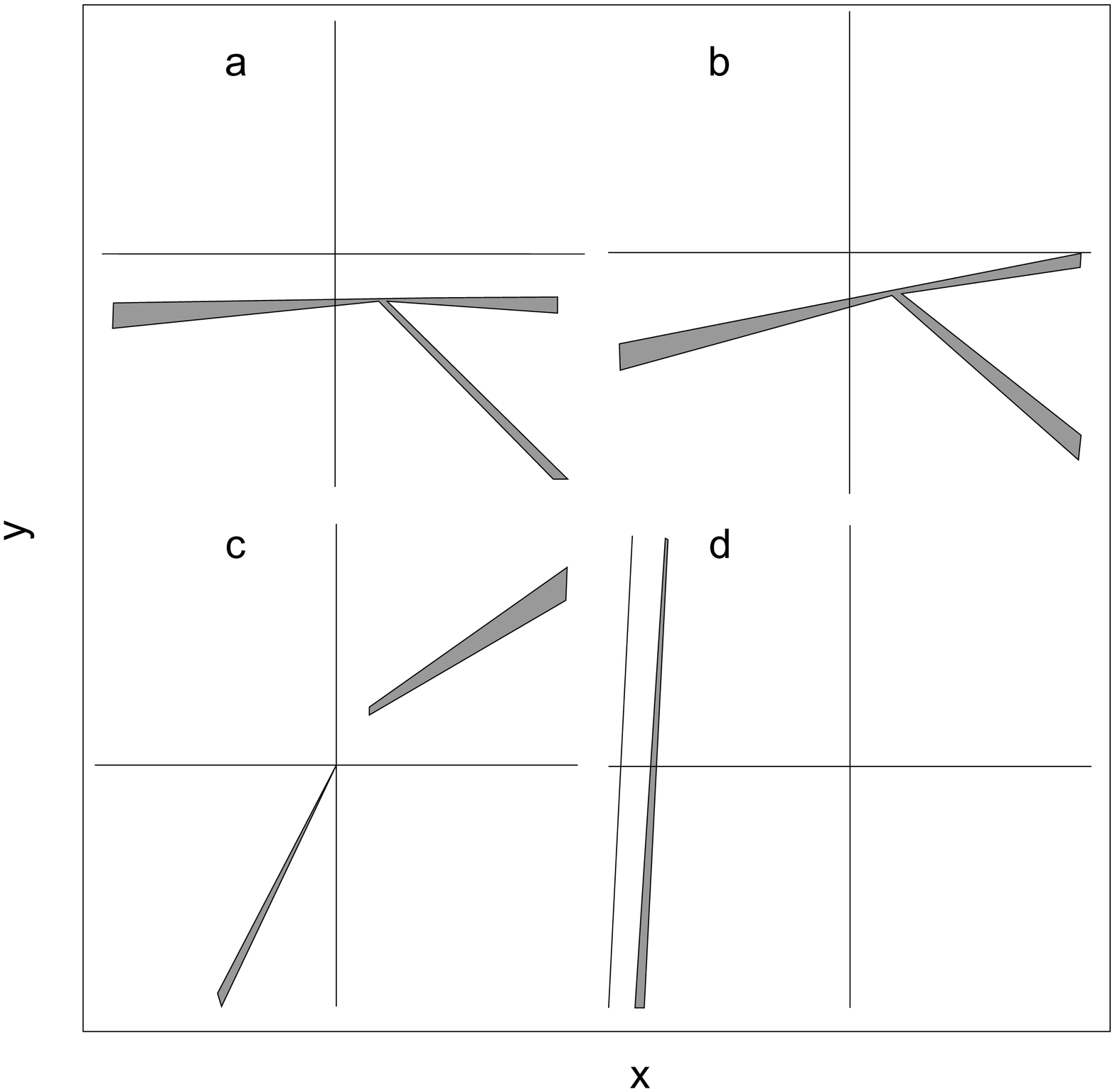}}
\hss} \caption{Some sets in the parameter space (filled with grey
color) which satisfy the energy constraint and provide the slow-roll
dynamics: $|\delta(\psi_i,\nu,\lambda)|<0.1$ and
$H^2(\psi_i,\nu,\lambda)>0$.} \label{Fig_istates}
\end{figure}

For simplicity we restrict numerical experiments  here by the
two-dimensional parameter domain $(\nu,\lambda)$ of non-minimal
couplings to gravity setting the  mass parameter $\mu$ zero. The
corresponding behavior of  $\delta(\psi)$ is shown on the Fig.
\ref{Fig_istates}. It appears that the sub-domains of the slow-roll
initial states are nearly one-dimensional in a wide range of
$\psi_i$. This  is true in the transplanckian region but changes
substantially when the field amplitude goes below the Planck scale.

\begin{figure}[p]
\hbox to\linewidth{\hss%
\psfrag{1}{\LARGE{$\psi_i=200$}}\psfrag{2}{\LARGE{$\psi_i=150$}}
\psfrag{3}{\LARGE{$\psi_i=100$}} \psfrag{4}{\LARGE{$\psi_i=50$}}
\psfrag{5}{\LARGE{$\nu=-\lambda=1$}}\psfrag{6}{\LARGE{$\nu=-\lambda=5$}}
\psfrag{7}{\LARGE{$\nu=-\lambda=10$}}\psfrag{8}{\LARGE{$\nu=-\lambda=15$}}

\psfrag{t}{\huge{$t$}} \psfrag{y}{\huge{$\psi,\,\alpha$}}

    \resizebox{10cm}{8cm}{\includegraphics{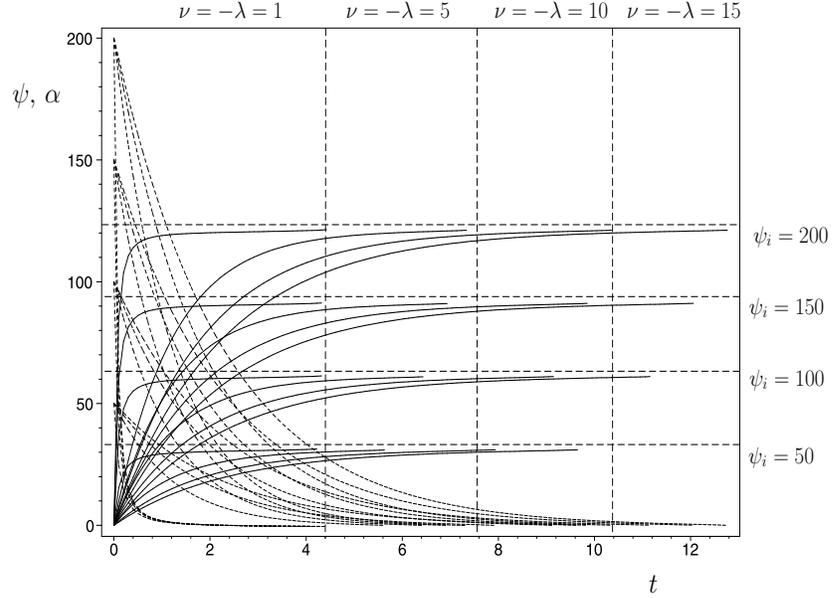}}
\hss} \caption{A set of 16 solutions for $\psi_i=50,100,150,200$
and $\nu=-\lambda=1,5,10,15$. Solid lines for the metric exponent,
$\alpha(t)$, dot lines for the YM field amplitude, $\psi(t)$.
Their linear similarity verifies that the trans-Planckian region
and the parameter line $\nu=-\lambda>1$ is a stable domain for the
slow-roll inflationary scenario.} \label{Fig_solutions}
\end{figure}

There is no practical sense in investigating such a complicated
domain of the initial states. So we then focus on the transplanckian
regime. The numerical solutions confirm that the line $\lambda=-\nu$
generates the stable slow-roll inflation in the area $\nu>1$. The
number of $e$-folds does not actually depend on the value of $\nu$,
and is proportional to $\psi_i$, as can be seen on the Fig.
\ref{Fig_solutions} . But it is very sensitive to the condition
$\lambda=-\nu$. Even the small deviations up to percent may increase
or decrease $\Ne$ up to several times. Finally, on the exit of the
inflationary stage when $\psi_e\sim 1$, the solutions shown on the
Fig. \ref{Fig_solutions}  meet the square-root singularity due to
the vanishing  system determinant $\Delta$. In fact, the singularity
can be avoided for  other choices of initial data, but the numerical
experiments performed here have shown that these usually do not
provide the good inflation.

It would be interesting to look whether there is any physics beyond
the condition $\lambda=-\nu$. In this case the non-minimal coupling
term in four dimensions will be $$\Lc=-\nu
[R_{\mu\nu}-(5/12)g_{\mu\nu}]A^{a\mu} A^{a\nu}.$$ The geometrical
tensor appearing here is close to, but not exactly coinciding with
the Einstein tensor. In such a model the stable inflationary regime
exists starting in the transplanckian region when the YM field
amplitude rolls down to the Planck  scale, while the number of
$e$-folds is proportional to the initial value of the field
amplitude, $\Ne\approx 0.6\psi_i$.

Another special property of the proposed non-minimally coupled YM
model is that typically there is no smooth exit from the
inflationary stage into the radiation dominated universe. The
solutions end with a square-root singularity, where the derivatives
$\dot\alpha$ and $\dot\psi$ diverge. In this area the system
demonstrates chaotic behavior due to the essentially non-linear
coupling of the YM field to the metric which then evolves from one
singularity to another. The solutions corresponding to radiation
dominated universe can be obtained when the initial energy of the YM
field is small, so that non-linear terms of the non-minimal coupling
can be neglected. Probably, the singular region between the
inflationary stage and the hot universe may correspond to a  phase
transition in the universe, when other matter fields are included.

\section*{Acknowledgments}
This work was supported by RFBR grants 11-02-01371-a and
11-02-01335-a.

\end{document}